\documentclass[prl,aps,twocolumn,showpacs,floatfix]{revtex4-1}

\usepackage{graphicx}
\usepackage{amsmath}
\usepackage{amssymb}
\usepackage{wasysym}

\newcommand{\bra}[1]{\langle #1|}
\newcommand{\ket}[1]{|#1\rangle}
\newcommand{\expectation}[1]{\langle #1\rangle}

\newcommand{\rhoss}{\rho_{\text{ss}}}
\newcommand{\rhoqq}{\rho_{q q_0}}
\newcommand{\Tr}{\mathop{\mathrm{Tr}}}
\newcommand{\tr}{\mathop{\mathrm{tr}}}
\newcommand{\Det}{\mathop{\mathrm{Det}}}
\newcommand{\none}{n^{(1)}_p}
\newcommand{\ntwo}{n^{(2)}_p} 
\newcommand{\nplus}{n^{(+)}_p}
\newcommand{\nminus}{n^{(-)}_p}

\begin{document}

\title{Dynamical Phases in the Full Counting Statistics of the Resonant-Level Model}

\author{Sam Genway}
\author{James M.~Hickey}
\author{Juan P.~Garrahan}
\author{Andrew D.~Armour}
\affiliation{School of Physics and Astronomy, University of Nottingham, Nottingham, NG7 2RD, United Kingdom}


\date{\today}

\begin{abstract} 
We present a thermodynamic formalism to study the full counting statistics (FCS) of charge transport through a quantum dot coupled to two leads in the resonant-level model.  We show that a close analogue of equilibrium phase transitions exists for the statistics of transferred charge;  by tuning an appropriate `counting field', crossovers to different \emph{dynamical} phases are possible.  Our description reveals a mapping between the FCS of a given device and current measurements over a range of devices with different dot-lead coupling strengths.  Further insight into features in the FCS is found by studying the occupation of the dot conditioned on the transported charge between the leads. 
\end{abstract}

\maketitle

\emph{Introduction---} The last two decades have seen significant interest in calculating and understanding the FCS of charge transport in mesoscopic systems~\cite{Levitov1996,*Nazarov2003, Blanter2000, Nazarov2003b, *Nazarov2003a, Flindt2005, Esposito2009, Flindt2009, Klich2003, Ivanov2010}.  Beyond the more readily accessible mean and variance, studying the FCS of transferred particles gives insights into coherences~\cite{Ferrini2009},  particle interactions~\cite{Kambly2011} and charge fractionalization~\cite{Carr2011}.  Various theoretical methods exist to extract FCS in mesoscopic systems: seminal works introduced an auxiliary system which is coupled to the mesoscopic system of interest during quantum transport~\cite{Levitov1996,Nazarov2003}.  An alternative theoretical approach is a two-point measurement scheme where the transferred charge is determined by the charge difference in a reservoir between two measurements separated in time~\cite{Schonhammer2007,Esposito2009,Doyon2011}.   This latter approach gives manifestly positive probabilities for the transferred charge and is minimally invasive since the coherent evolution of the system is not interrupted or altered between the two measurements.  Over the last few years, experiments have provided increasingly detailed measurements of FCS.  For example, a quantum point contact has been used to measure the sequential tunnelling of electrons through a quantum dot in real time~\cite{Gustavsson2006,Flindt2009}.
  

Recent work has developed an ``$s$-ensemble'' approach~\cite{Garrahan2007, *Hedges2009a} for studying the counting statistics of open quantum systems~\cite{Garrahan2010} which is based on the application of large-deviation methods~\cite{Touchette2009}.  This approach allows an analogy to be developed between quantum trajectories and traditional thermodynamics.  This large-deviation method has been applied to the time record of quantum jumps obtained by unravelling a Markovian master equation (MME) and has revealed new features in the dynamics of open quantum systems~\cite{Garrahan2011, *Genway2012, *Ates2012}.  Most notable is the existence of $\emph{dynamical}$ phase transitions between phases with different quantum-jump rates.  These transitions may be crossed by tuning system parameters or by adjusting a counting field which biases the system towards rare trajectories where the number of quantum jumps differs from the mean.  This method has also been applied to transport problems approximated by a MME~\cite{Li2011}.

In this Letter, we introduce a new formulation of the $s$-ensemble which captures features beyond those accessible with a MME description of the dynamics.  We use the two-point measurement approach to study quantum transport, where we consider a trajectory to be specified only by the charge transferred between two measurements separated by a time $t$.  This allows us to study the statistics of transport without making Born and Markov approximations.  We apply the method to an exactly solvable model for a quantum dot coupled to two free-electron leads and find a rich dynamical phase diagram not captured in the MME approach~\cite{Li2011}.   The approach uncovers a mapping relating the FCS of a system to the average current in a class of related systems.  This remarkable property of the model is not revealed in other approaches to FCS.  

\begin{figure}
\includegraphics[width=7cm]{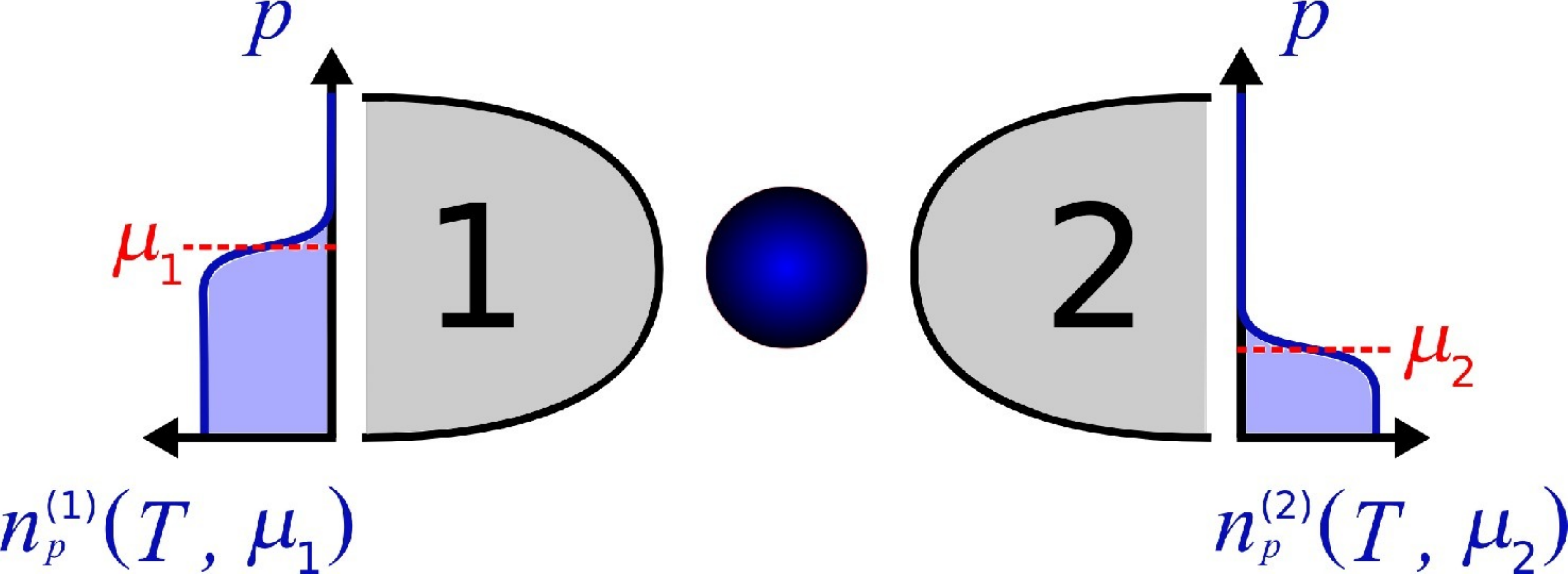}
\caption{A schematic diagram of the quantum dot and semi-infinite conducting leads 1 and 2 with thermal occupations $n_p^{(1)}$ and $n_p^{(2)}$ at temperature $T$ and chemical potentials $\mu_1$ and $\mu_2$ in the non-equilibrium steady state.}
\label{dot}
\end{figure}

\emph{Transport model ---} We study a single-level quantum dot coupled to two semi-infinite conductors held at different chemical potentials, shown schematically in Fig.~\ref{dot}.  The universal low-energy behavior of free electrons scattering from a single dot level is captured by the resonant-level model~\cite{Doyon2011, Doyon2006, Anderson1961, Wiegmann1983}.  Using this model and the two-point measurement method, Bernard and Doyon~\cite{Doyon2011} recently provided an exact derivation of the characteristic function for charges transferred through the dot after long times, and demonstrated its equivalence to the established result obtained by Levitov and coworkers~\cite{Levitov1996}.
The steady-state distribution of charge transferred through the dot $\Delta q$ during a time $t$ takes a large deviation (LD) form $P_t(\Delta q) = e^{-t\phi(\Delta q/t)}$~\cite{Touchette2009}.  Therefore, the moment generating function $Z_t(s)$, found via Laplace transform, also takes a LD form $Z_t(s) = \sum_{\Delta q} e^{-s\Delta q} P_t(\Delta q) = e^{t \theta(s)}$.  

This mathematical framework allows a thermodynamic analogy to be made where the convex LD functions $\phi(\Delta q/t)$ and $\theta(s)$ are considered akin to, respectively, entropy and free energy densities~\cite{Garrahan2007,Garrahan2010}.  $Z_t(s)$ plays the r\^ole of a \emph{dynamical} partition function, while the parameter $s$ can be considered a \emph{time}-intensive conjugate field to the \emph{time}-extensive number of transferred charges $\Delta q$.  Tuning this `$s$-field' allows us to bias the system towards rare occurrences where $\Delta q$ is greater ($s<0$) or smaller ($s>0$) than the average.  We identify \emph{dynamical} phase transitions with singular regions in $\theta(s)$ and establish an order parameter, the average current $I(s) = \expectation{\Delta q}/t$, as a way to distinguish dynamical phases. Through its $s$-derivatives, $I(s)$ contains the full set of cumulants for charge transport and hence provides a description of the FCS of the system.

After linearising the spectrum about the Fermi energy and unfolding the model, we consider the Hamiltonian~\cite{Doyon2006,Anderson1961,Wiegmann1983}
\begin{equation}
H = -i \sum_{j=1}^2 \int_{-L}^L \!\!\!\!dx \psi_j^\dag \partial_x \psi_j + \frac{\tau}{\sqrt{2}} \left(\psi_j^\dag(0) d + d^\dag \psi_j(0)\right) + \epsilon d^\dag\! d.
\label{eq:H}
\end{equation}
Here, $d^\dag$ ($d$) is the fermionic creation (annihilation) operator for the dot at energy $\epsilon$ and $\psi_j^\dag(x)$ ($\psi_j(x)$), with $j=1,2$ the creation (annihilation) operators for the unfolded leads 1 and 2 at position $x$.  The leads have length $L$, which we will take to infinity, and couple to the dot with tunnelling matrix element $\tau$.  Following convention~\cite{Wiegmann1983}, we have set the energy scale of the problem by choosing unit Fermi velocity, with density of states at the Fermi energy $g=1/2\pi$.  


In the limit $L\longrightarrow\infty$, the Hamiltonian~\eqref{eq:H} is diagonalised with even (odd) mode operators $a^\dag_p$ ($b^\dag_p$) with eigenenergy $p$ such that $H= \int p (a^\dag_p a_p + b^\dag_p b_p) dp$.  The field operators are then expressed 
\begin{eqnarray}
\psi_1(x,t)+\psi_2(x,t) &=& \int \frac{dp}{\sqrt{\pi}} e_p(x) e^{ip(x-t)} a_p\nonumber \\
\psi_1(x,t)-\psi_2(x,t) &=& \int \frac{dp}{\sqrt {\pi}} e^{ip(x-t)} b_p \nonumber\\ 
d(t) &=& \frac{i}{\tau}\int \frac{dp}{\sqrt{2\pi}} w_p e^{-ipt} a_p
\label{eq:modes}
\end{eqnarray}
in the Heisenberg representation.  The even mode hybridises with the dot level such that $w_p = \tau^2/[-\tau^2+i(p-\epsilon)]$ and a phase shift is introduced into the even mode due to scattering off the dot, such that $e_p=e^{i \delta_p} = 1-w_p$ when $x>0$ and $e_p=1$ for $x<0$.  The transmission coefficient through the dot for electrons with energy $p$ is thus given by 
\begin{equation}
T_p = |\sin(\delta_p/2)|^2 = \frac{\tau^4}{\tau^4 + 4(p-\epsilon)^2}.
\label{eq:trans} 
\end{equation}

We consider the case where the system has a steady-state density matrix $\rhoss$~\cite{Doyon2011,Hershfield1993} with different chemical potentials $\mu_i$ in leads $i=1$ and 2 (see Fig.~\ref{dot}).  We find the transferred charge by projecting on to charge states of lead 1 at times separated by $t$:  after projecting on to states of lead 1 with charges $q_0$ and $q$ at times 0 and $t$, the density matrix takes the form $\rhoqq = P_q U_t P_{q_0} \rhoss P_{q_0} U^\dag_t P_q$, with $U_t = e^{-iHt}$ the evolution operator and $P_q$ the charge projection operator on lead 1.  The probability of measuring charge $q_0$ in the first measurement and then charge $q$ after time $t$ is thus $\Tr(\rhoqq)$ and the LD function for the transferred charge, $\Delta q = q_0-q$, is
\begin{equation}
\theta(s)=t^{-1}\log\sum_{q q_0}e^{-s(q_0-q)}\Tr(\rhoqq)\,.
\label{eq:thetadef}
\end{equation}
From this, we extract the order parameter, the average current $I(s) = -\theta'(s)$, where the prime denotes $s$-differentiation.  We also study the charge fluctuations $\Delta I^2(s) = [\expectation{\Delta q^2}-\expectation{\Delta q}^2]/t = \theta''(s)$.  

Following~\cite{Doyon2011}, $\theta(s)$, from its definition in Eq.~\eqref{eq:thetadef}, is
\begin{multline}
\theta(s) = \int \frac{dp}{2\pi} \log\Big\{1+T_p\big[\none(\ntwo-1)(1-e^{s})\\
+\ntwo(\none-1)(1-e^{-s}) \big]\Big\}\,,
\label{eq:theta} 
\end{multline}
where $\none$ ($\ntwo$) is the thermal mode occupation at energy $p$ for lead 1 (2), shown in Fig.~\ref{dot}.  This result matches the Levitov formula~\cite{Levitov1996}, as proven in~\cite{Doyon2011}.  The $s$-ensemble approach also provides a convenient way of investigating the state of the dot conditioned on the transported charge by finding the dot occupation for rare charge trajectories.  We consider the dot occupation $n_d(s)$ in the $s$-biased distribution of measured charges,
 \begin{equation}
 n_d(s)= \frac{1}{Z_t(s)}\sum_{q q_0} e^{-s(q_0-q)} \Tr(d^\dag \!d\,\rhoqq)\,,
 \label{eq:Os}
 \end{equation}
 which can be evaluated analytically for the resonant-level model (see Supplemental Material~\cite{supp}).

\begin{figure}
\includegraphics[width=8.6cm]{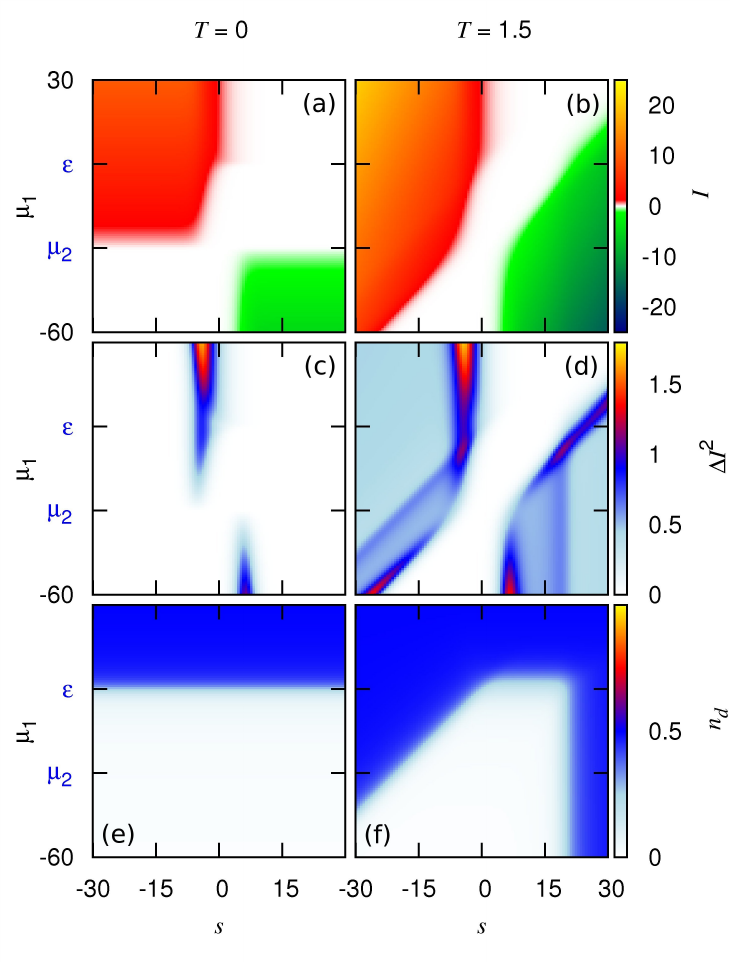}
\caption{(Color online.) Plots of (a,b) the current order parameter $I(s)$, (c,d) current fluctuations $\Delta I^2(s)$, and (e,f) the dot occupation probability $n_d(s)$ for $T=0$ (a,c,e) and $T=1.5$ (b,d,f) as a function of $s$ and $\mu_1$.  We set $\mu_2=-30$ and $\epsilon=0$ as labelled on the ordinate axis.}
\label{multiphase}
\end{figure}

\emph{Dynamical Phases ---} We now discuss the statistics of transferred charge in the $s$-ensemble picture.  To be concrete, we fix $\mu_2=-30$ below the dot energy and study the charge dynamics with different $s$-fields as $\mu_1$ is changed.  A summary of our results is shown in Fig.~\ref{multiphase} where we set $\tau=2$ and, without loss of generality, choose the dot energy $\epsilon=0$.  

We first examine the zero-temperature dynamical phase diagram.  Using the mean current $I(s) = -\theta'(s)$ as an order parameter to distinguish different regimes, we find three distinct `phases' separated by sharp crossovers in both $s$ and $\mu_1$.  Figure~\ref{multiphase}(a) shows distinct regions where charge flows from left to right ($I(s)>0$), no charge flows ($I(s)\simeq 0$) and charge flows from right to left ($I(s)<0$).  Analogous to equilibrium phase transitions, Fig~\ref{multiphase}(c) shows that crossover lines between dynamical phases are marked by sharp peaks in current fluctuations $\Delta I^2(s)$, except where these lines are parallel to the $s$-axis.  The phases are divided by distinct crossovers which become sharper with increasing $\mu_1-\mu_2$.
However, when $\mu_1 = \mu_2$, not only is the \emph{average} current zero, there are no fluctuations in transferred charge.  Therefore, application of the $s$-field is unable to modify the distribution of charge trajectories and $I(s)=0$ everywhere.  Upon raising $\mu_1$ above $\mu_2$, charges can flow from the left lead to the right.  We now see that applying the $s$-field allows new dynamical phases to be reached by biasing the mean current to larger values if $s<0$ and suppressing the current for $s>0$. 

The dynamical phase structure is modified at finite temperature as there is always a finite probability of charge flowing in either direction.  Figure~\ref{multiphase}(b) illustrates the case of $T=1.5$.  When $\mu_1=\mu_2$, the system is in equilibrium with $I(0)=0$.  However, at finite $T$, there are fluctuations in the charge transported between measurements and application of the $s$-field takes $I(s)$ through sharp crossovers even when $\mu_1=\mu_2$.  As $\mu_1$ is increased, dynamical phases with positive and negative $I(s)$ are still present, but with increasing forward bias the $s$-field required to reach the negative current regime grows in proportion to $\mu_1/T$ because transport against the bias becomes increasingly rare.  The phase crossovers are again marked by peaks in the fluctuations $\Delta I^2(s)$, but we now see finite fluctuations within the phases with with large $|I(s)|$ (see Fig.~\ref{multiphase}(d)).  Within these phases, $\Delta I^2(s)$ is virtually independent of $s$ and the potential bias and is proportional to $T$.  There are smaller peaks in $\Delta I^2(s)$ in the low-bias regime $\mu_1<\epsilon$ which are additional to those at the phase boundaries.  We return to these features later.

The dynamics of the system are constrained by a fluctuation theorem (FT) for the rate of entropy production~\cite{Esposito2009}, and this imposes a structure on the dynamical phase diagram.  The probabilites of transporting charge in the forward and reverse directions are related by ${P_t(\Delta q)}/{P_t(-\Delta q)} = e^{2 s_{I=0}}$.  From Eq.~\eqref{eq:theta}, $s_{I=0} = (\mu_1-\mu_2)/2T$ is the $s$-field required to bring the current to zero.   As a consequence of the FT, $\theta(s)$ is symmetric about $s_{I=0}$ such that $I(s-s_{I=0}) = -I(s_{I=0}-s)$.  The form of $s_{I=0}$ illustrates that the $s$-field shares characteristics with a real potential bias: at finite $T$ it can be tuned to oppose the bias $\mu_1-\mu_2$ and bring the system to equilibrium.

\emph{FCS from current measurements---} A remarkable feature of the dynamical phases at $T=0$ is that rare charge trajectories, where $s\ne 0$, capture the typical charge trajectories for all dot-lead coupling stengths $\tau$.  Specifically, from Eq.~\eqref{eq:theta}, 
\begin{equation}
I(s) = \int_{\mu_2}^{\mu_1} \frac{dp}{2\pi}\, \frac{e^{-s} \tau^4}{4p^2+e^{-s} \tau^4}\,,
\label{eq:Iofs}
\end{equation}
hence there exists a simple mapping between the behavior at a given $s$ and that of the unbiased ($s=0$) dynamics by modifying the coupling $\tau\rightarrow\tau_s=e^{-s/4}\tau$.  The phase diagram for another coupling $\tau'$ is thus straightforwardly related to the phase diagram at $\tau$ by a translation along the $s$-axis where $s\rightarrow s-4\zeta\log(\tau'/\tau)$, with $\zeta=\text{sgn}(\mu_1-\mu_2)$.   Furthermore, since $I(s)$ encodes all the cumulants of the current through its derivatives, the FCS for a particular $\tau$ can be inferred from the measurements of the average current for a range of coupling strengths.

\begin{figure}
\includegraphics[width=8cm]{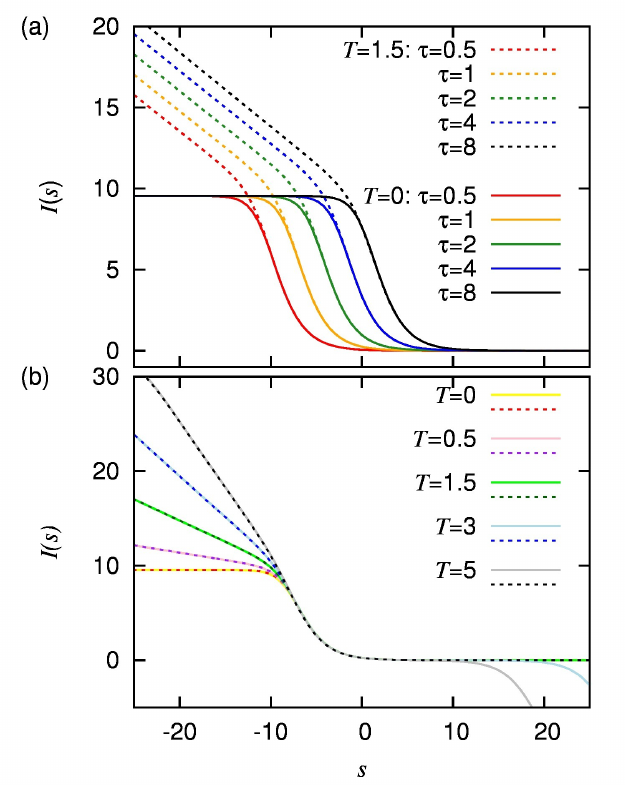}
\caption{(Color online.) Plots of $I(s)$ for $\mu_1=-\mu_2=30$. (a) shows the translational property of $I(s)$ with $\tau$, at zero and finite $T$, as labelled.  (b) demonstrates $I(s)$ (solid lines) can be found from the unbiased current (dashed lines) through quantum dots with the modified $\tau\rightarrow\tau_s$ and $\mu_{1,2}\rightarrow\mu_{1,2}^s$ (see main text), for $\tau = 1$ and different $T$, as labelled.}
\label{FCS}
\end{figure}

More generally, at finite $T$, we find that FCS of a system with an $s$-bias applied are replicated in the $s=0$ FCS for another quantum-dot system.  In general, energy-dependent couplings must be introduced to the model (see Supplemental Material~\cite{supp}).  However, provided $|\mu_{1,2}|/T \gg 1$, with ${\mu_2}<\epsilon<{\mu_1}$, modifications to the form of the Hamiltonian~\eqref{eq:H} are not needed.  The translation property of $I(s)$ still holds, as shown in Fig.~\ref{FCS}(a).  Furthermore, after introducing the same modified coupling $\tau \rightarrow \tau_s$ and mapping $\mu_{1,2} \rightarrow \mu_{1,2}^s$, where $\mu_{1,2}^s$ satisfies
\begin{equation}
\mu_{1,2}^s = \mu_{1,2} \pm T \log \frac{4 (\mu_{1,2}^s)^2 + \tau_s^4}{4 (\mu_{1,2}^s)^2 + 1}\,,
\label{eq:mus}
\end{equation}
we find that we can determine $I(s)$ from values of the unbiased current, $I(0)$, in resonant-level systems with coupling $\tau_s$ and chemical potentials $\mu_{1,2}^s$.  We demonstrate in Fig.~\ref{FCS}(b) that good agreement exists between this scheme and the exact $I(s)$ for a range of temperatures when $s<s_{I=0}$.  $I(s>s_{I=0})$ can be inferred using the reflection antisymmetry in $I(s)$ described by the FT.  
This simple structure for the mapping no longer holds when the bias is very small as additional features in the FCS emerge.  These manifest as additional peaks in the fluctuations $\Delta I^2(s)$ (see Fig.~\ref{multiphase}(d)). Unlike the peaks in $\Delta I^2(s)$ associated with crossovers in $I(s)$, the width and height of these secondary peaks are $\tau$-dependent.

\emph{Rare trajectories and dot occupation---} We now return to the regime with $\mu_{1,2} <\epsilon$ and consider the additional peaks in $\Delta I^2(s)$, which occur along the lines $\mu_1 = sT$ and $s = -\mu_2/T$ (see Fig.~\ref{multiphase}(d)).  While these features necessarily originate from the form of the LD function $\theta(s)$, or the dynamical order parameter $I(s)$, a physical explanation of their existence is not obvious.  However, we gain insight by studying the state of the dot as a function of the $s$-bias, $n_d(s)$, given in Eq.~\eqref{eq:Os}.  As shown in Fig.~\ref{multiphase}(f), $n_d(s)$ shows a marked crossover from zero to 0.5 which coincides with these peaks in $\Delta I^2(s)$, showing that the dot occupation becomes correlated with features in the FCS.  This crossover in dot occupation results from a crossover in the tunnelling rates on and off the dot as $s$ and $\mu_{1,2}$ are tuned.
In Fig.~\ref{multiphase}(e) we see at the dot occupation $n_d(s)$ is $s$-independent when $T=0$, and therefore does not give rise to features in $\Delta I^2(s)$.

\emph{Conclusions---}  We have presented a new approach to the study of FCS based on a thermodynamic formalism, taking the fully-coherent dynamics of the resonant-level model as an exactly-solvable example.  We identified close analogues of equilibrium phases with sharp crossovers between dynamical phases marked by large peaks in the fluctuations of transported charge.  The $s$-biased current $I(s)$ contains all cumulants in the FCS through its $s$-derivatives.  However, studying FCS this way reveals a remarkable mapping between the FCS of quantum-dot systems across parameter space.  We further show that the FCS may be mapped to the current through quantum dots at several values of the dot-lead coupling and applied bias.  Features in the current fluctuations become correlated not only with the behavior of the current, but also with the occupation of the dot in rare trajectories, $n_d(s)$. 
At finite bias, the dynamical phases are marked by sharp crossovers rather than transitions.  However, we anticipate that transport systems with more complex internal dynamics will show sharp dynamical transitions in this space of rare trajectories.  We believe that analysis of more complex systems with the approach developed here, for example those with interactions~\cite{Carr2011} or dissipation~\cite{Chung2009}, will prove fruitful.

\emph{Acknowledgements---} We wish to thank Adrian Budini for insightful discussions.  We are grateful for financial support from The Leverhulme Trust under Grant No.~F/00114/B6.

\newpage

\phantom{.}
\newpage
\pagestyle{empty}
\begin{center}
\emph{\bf{SUPPLEMENTARY MATERIAL}}

for

{Dynamical Phases in the Full Counting Statistics of the Resonant-Level Model}
\end{center}

In this supplementary material, we sketch the derivations of $\theta(s)$ and $n_d(s)$ for the resonant-level model and identify the mapping of $s\ne 0$ statistics in the resonant-level model to the unbiased statistics of a more general class of models.
\newline

\emph{Derivation of $\theta(s)$ and $n_d(s)$ ---} From the definition of $\theta(s)$ in Eq.~\eqref{eq:thetadef}, the derivation of Eq.~\eqref{eq:theta} follows Bernard and Doyon~\cite{Doyon2011} but with a rotation in the complex $s$-plane.  We sketch the details as we use these as a foundation for our derivation of the dot occupation $n_d(s)$.  We begin by expressing the dynamical partition function as

\begin{equation}
Z_t(s) = \frac{1}{2\pi}\Tr\left(\rhoss \int_0^{2 \pi}\!\! d\eta \,\, e^{F(s,t,\eta)}\right)
\end{equation}
 where
 \begin{equation}
 e^{F(s,t,\eta)} = e^{(s/2 - i\eta)Q} e^{-s U^\dag_t Q U_t} e^{(s/2+i \eta)Q}\,.
 \end{equation}
The dynamical partition function may be evaluated using a formula by Klich~\cite{Klich2003} $\Tr(e^A e^B e^C) = \Det(1+e^a e^b e^c)$ where $A, B, C$ are operators bilinear in fermion operators and $a, b, c$ are associated single-particle operators.  Constructing single-particle matrix elements as $2\times 2$ blocks in the even and odd mode operators, with single-particle states of energy $p$ denoted by $\ket{p} = (a^\dag_p \ket{\text{vac}}, b^\dag_p \ket{\text{vac}})^T$, one expresses the matrix elements $\bra{p}e^{f(s,t,\eta)}-1\ket{p'} = M(p,\eta) (e^{i(p-p')t} -1)/ip$, neglecting exponentially decaying terms when $t \gg g v_F \tau^{-2}$.  Here, $f$ is the single-particle operator corresponding to $F$ and 
\begin{multline}
 M(p,\eta) = -2 \sinh \frac{s}{2} \bigg[\sqrt{T_p(T_p-1)}(\sigma_y\cos\eta  +\sigma_z\sin\eta)\\
  - T_p\left(\sigma_x\cosh\frac{s}{2} + I\,\sinh\frac{s}{2} \right)\bigg]\,.
  \label{eq:F}
\end{multline}
Here, $\sigma_{x,y,z}$ are the Pauli matrices with $I$ the identity.  From the Klich formula we then have 
\begin{eqnarray}
Z_t(s) &=& \int_0^{2\pi} \frac{d\eta}{2\pi} \frac{\Det[I+ r_{ss}e^{f(s,t,\eta)}]}{\Det[I + r_{ss}]}\nonumber\\ 
&=& \int_0^{2\pi}\frac{d\eta}{2\pi}\Det[I + N (e^{f(s,t,\eta)}-1)]
\end{eqnarray}
where $r_{ss}$ is the single-particle operator corresponding to the Hershfield density matrix $\rhoss$ and $N = (I+r_{ss})^{-1}r_{ss}$ which is diagonal in $p$ with elements $N_p=\bra{p}N\ket{p}=\nplus I + \nminus \sigma_x$.  With careful evaluation of $\log Z_t(s)$, one ultimately finds this to be linear in $t$ with contributions only from diagonal blocks in $p$~\cite{Doyon2011} so that Eq.~\eqref{eq:theta} is found.

We now turn to the derivation of $n_d(s)$, from the definition in Eq.~\eqref{eq:Os}, which is plotted in Fig.~\ref{multiphase}(e) and (f).  Since $d^\dag \!d$ commutes with projection operators on the Hilbert space of the leads, we find
\begin{equation}
n_d(s) = \frac{1}{Z_t(s)}\frac{\partial}{\partial j}\Tr\left(\rhoss \,\, \int_0^{2\pi}\!\!\frac{d\eta}{2\pi}\,\, e^{F(s,t,\eta)} e^{\,j U^\dag_t d^\dag\! d U_t} \right)\bigg|_{j=0}
\end{equation}
where our construction allows the Klich formula to be used.  Thus
\begin{equation}
n_d(s) = \int_0^{2\pi}\frac{d\eta}{2\pi} \frac{\partial}{\partial j} \log \Det\left[I+ N e^{f(s,t,\eta)} e^{D(t)}\right]\Big|_{j=0}
\label{eq:ndSUPP}
\end{equation}
where $e^{D(t)}$ is the single-particle version of $e^{\,j U^\dag_t d^\dag\! d U_t}$.  Now, for brevity, denoting $M(p,\eta)=M_p$ and using $\delta_t(p-p')=(e^{i(p-p')t} -1)/{i(p-p')}$, we find
\begin{multline}
\bra{p}e^{f(s,t,\eta)} e^{jD(t)}-1\ket{p'} = \\
M_p\delta_t(p-p') + (e^j-1) J_{pp'} e^{i(p-p')t}+\\
 (e^j-1)M_p J_{pp'} e^{i(p-p')t}\delta_t(p-p')
\end{multline}
where $J_{pp'} = \tau^{-2} w_p w_{p'}^* (I+\sigma_z)/2$.  Using this and expanding~\eqref{eq:ndSUPP} we find
\begin{multline}
n_d(s) = \int_0^{2\pi}\frac{d\eta}{2\pi}\frac{\partial}{\partial j} \sum_{k=1}^\infty \frac{(-1)^{k+1}}{k} \Tr [N_p (M_p \delta_t(p-p')\\
 +  (e^j-1)J_{pp'}e^{i(p-p')t} +  (e^j-1)M_p J_{pp'}\delta_t(p-p')e^{i(p-p')t}) ]^k\,.
  \label{eq:toprove}
 \end{multline}
Equation~\eqref{eq:toprove} proves to be time independent.  To see this, we use the identity $\delta_t(p) = \int_0^t dt' e^{ipt'}$ from which we see the $k^\text{th}$ term in the sum after $j$-differentiation has the form
\begin{multline}
n_d(s) = \int dt_1\ldots dt_{k-1} \int dp_1 \ldots dp_k \\
\sum_{l=1}^k \prod_{m=1}^k e^{i p_m(t_m-t_{m-1})} e^{-i(p_l-p_{l+1})(t_l-t)}\\
\left[e^{i\sum_{j=1}^k \alpha_j p_j} + e^{i\sum_{j=1}^k \beta_j p_j} \delta_t(p_l-p_{l+1})\right]
\end{multline}
where we consider, instead of the specific $p_j$-dependent matrices, general distributions over $\alpha_j$ and $\beta_j$ as in~\cite{Doyon2011} and omit the $\eta$-integration.  Repeatedly using the Fourier representation of the delta function, we find the above expression collapses to $k\int dp \left[ e^{i(\sum_{j=1}^k \alpha_j) p} +  e^{i(\sum_{j=1}^k \beta_j) p}\right]$.  Therefore Eq.~\eqref{eq:toprove} takes the explicitly time-independent form
 \begin{multline}
 n_d(s)=\int_0^{2\pi}\frac{d\eta}{2\pi} \sum_{k=1}^\infty (-1)^{k+1} \\ \int dp \tr [(M_p N_p)^{k-1}(I+M_p)J_{pp}N_p]
\end{multline}
with $\tr$ denoting the trace over $2\times 2$ blocks.  After re-summing the series, we find 
 \begin{widetext}
\begin{equation}
n_d(s) =  \int_0^{2\pi}\frac{d\eta}{2\pi} \int dp \tr[(I+M_p N_p)^{-1} (I+M_p)J_{pp}N_p] 
\end{equation} 
which, upon substituting the $2\times 2$ matrices into the integrand and performing the $\eta$-integral yields 
\begin{equation}
n_d(s) = \int \frac{dp}{2\pi}\,\frac{|w_p|^2}{2}\left[1- \frac{2\nplus-1}{(\nplus-2 \none \ntwo)T_p(1-\cosh s) + 2 \nminus T_p \sinh s  -1} \right] 
\label{eq:nd2}
\end{equation}
\end{widetext}
where $n_p^{(\pm)} = n_p^{(1)}\pm n_p^{(2)}$.
 \newline

\emph{Mapping $s\ne 0$ statistics at finite temperature to $s=0$ statistics ---}
In the main text of the paper, we highlight how in the zero temperature limit the rare, \emph{i.e.} $s\ne 0$, charge trajectories of one system are the typical, \emph{i.e.} $s=0$, trajectories of another.  This can be seen by a simple rescaling of the coupling $\tau$.  At non-zero temperature, the order parameter $I(s)$ is still an integral over all lead modes $I(s) = \int I_p(s) dp/2\pi$ as in Eq.~\eqref{eq:Iofs}, but the integrand $I_p(s)$ is modified by $s$ and takes the form
\begin{equation}
\frac{[e^{-s} \none(\ntwo-1)+e^s \ntwo(1-\none)] T_p}{[e^{-s} \none(\ntwo\!\!-\!\!1) \!+\! e^s (\none\!\!-\!\!1)\ntwo\!\!+\!(2\nplus\!\!-\!\!2 \none \ntwo) ]T_p\!-\!1}
\label{eq:integrand}
\end{equation}
From this expression and its derivatives we find all the cumulants of the $s$-biased system.  These are the same as those of an unbiased system with lead occupations $\tilde{n}_p^{(1)}$ and $\tilde{n}_p^{(2)}$ and effective transmission coefficients $\tilde{T}_p$.  We find
\begin{equation}
\tilde{n}_p^{(1)} = \frac{\none e^{-s}}{1+\none(e^{-s}-1)}
\end{equation}
and
\begin{equation}
\tilde{n}_p^{(2)} = \frac{\ntwo e^{s}}{1+\ntwo(e^{s}-1)}
\end{equation}
which amounts to shifts in the chemical potentials in the two leads where $\mu_1 \rightarrow \mu_1 + sT$ and $\mu_2 \rightarrow \mu_2 - sT$.  (This is consistent with our understanding that the $s$-bias on the statistics has characteristics of a real potential bias; this is discussed in the main text with reference to the $s$-bias $s_{I=0}$ required to bring $I(s)$ to precisely zero.)  The transmission coefficients must also be modified in a way that is more involved than at $T=0$.  The new transmission coefficients $\tilde{T}_p$ take the form of Eq.~\eqref{eq:trans} but with $p$-dependent parameters $\tau \rightarrow \tilde{\tau}_p = \tau \chi_p$ with
\begin{equation}
\chi_p^4 = \left[1+(e^{-s}-1)\none \right]\left[1+(e^{s}-1)\ntwo \right]\,.
\end{equation}
We note that $\tilde{T}_p=T_p$ when $T_p=0$ or 1 and, for any real $s$, $0\le T_p \le 1$ implies that $0\le \tilde{T}_p \le 1$.  

In general, the transmission coefficients take a new $p$-dependent form which cannot be captured by a model with energy-independent couplings between the dot and the leads.  However, discussed in the main text, an approximate mapping exists when  $|\mu_{1,2}|/T \gg 1$, with $\mu_2<0<\mu_1$ ($\epsilon=0$), which avoids energy-dependent couplings.
As above, the form of $I(s)$ is an integral over all $p$-modes.  However, where $p \gg \mu_2$ the integrand $I_p(s)$ is approximately
\begin{equation}
\frac{\tau_s^4}{4(e^{(p-\mu_1)/T}+1)p^2 + (e^{(p-\mu_1)/T+s} +1)\tau_s^4 }
\end{equation}
with $\tau_s = e^{-s/4}\tau$ as in the main text.  It is clear that this expression that the form of the $s=0$ integrand, but with $\tau\rightarrow\tau_s$ and an effective modification to the chemical potential $\mu_1$ in the second term in the denominator.  An equivalent expression exists involving only the lower chemical potential $\mu_2$ when $p\ll \mu_1$:
\begin{equation}
\frac{\tau_s^4}{4(e^{-(p-\mu_2)/T}+1)p^2 + (e^{-(p-\mu_2)/T-s} +1)\tau_s^4 }
\end{equation}
However, provided $|\mu_{1,2}|/T \gg 1$, when $\mu_2 \ll p \ll \mu_1$, we find the simple $\mu_{1,2}$-independent form $I_p(s) = \tau_s^4/(4p^2+\tau_s^4)$.  Therefore we are able to find a new system, with modified parameters $\tau'$ and $\mu'_{1,2}$, such that $I_p(0)$ for this new model is almost exactly equal to $I_p(s)$ for the first system.  We do this by matching the regimes $p\ll \mu_1$ and $p \gg \mu_2$ which, due to the $\mu_{1,2}$-independent form for intermediate $p$, leads to $I_p(0)$ for the new system matching $I_p(s)$ for all $p$.  For example, at large $p$ we require
\begin{multline}
\frac{\tau_s^4}{4(e^{(p-\mu_1)/T}+1)p^2 + (e^{(p-\mu_1)/T+s} +1)\tau_s^4 } \\
\simeq \frac{\tau'^4}{4(e^{(p-\mu'_1)/T}+1)p^2 + (e^{(p-\mu'_1)/T} +1)\tau'^4 }
\label{eq:fitting}
\end{multline}
from which we require the new system to have $\tau' = \tau_s$ and upper chemical potential given by $\mu'_1 = \mu_1^s$ in Eq.~\eqref{eq:mus}.  $\mu'_2$ is found by an equivalent condition, with the agreement becoming exact in the limit $T \rightarrow 0$ (see Fig.~3(b) in the main text).


\end{document}